\documentclass[conference]{IEEEtran}
\IEEEoverridecommandlockouts

\usepackage{cite}
\usepackage{amsmath,amssymb,amsfonts}
\usepackage{algorithmic}
\usepackage{graphicx}
\usepackage{textcomp}
\usepackage{xcolor}
\def\BibTeX{{\rm B\kern-.05em{\sc i\kern-.025em b}\kern-.08em
    T\kern-.1667em\lower.7ex\hbox{E}\kern-.125emX}}
\begin{document}

\title{MLC-Agent: Cognitive Model based on Memory-Learning Collaboration in LLM Empowered Agent Simulation Environment\\
}

\author{
\begin{tabular}{cc}
\parbox{7cm}{
  \centering
  \textbf{Ming Zhang\textsuperscript{*}\thanks{This work has been supported in part by National Key Research and Development Program of China (No.2021YFF0900800), National Natural Science Foundation of China (No. 62472306, No. 62441221, No. 62206116), Tianjin University's 2024 Special Project on Disciplinary Development (No. XKJS-2024-5-9), Tianjin University Talent Innovation Reward Program for Literature \& Science Graduate Student (C1-2022-010), and Shanxi Province Social Science Foundation (No.2020F002).}}\\
  \textit{Faculty of Environment, Science and Economy}\\
  \textit{University of Exeter}\\
  Exeter, UK\\
  \texttt{mz9408@outlook.com}
}
&
\parbox{7cm}{
  \centering
  \textbf{Yiling Xuan\textsuperscript{*}}\\
  \textit{College of Intelligence and Computing}\\
  \textit{Tianjin University}\\
  Tianjin, China\\
  \texttt{yiling\_xuan@163.com}
} \\[0.8cm]
\parbox{7cm}{
  \centering
  \textbf{Qun Ma}\\
  \textit{College of Intelligence and Computing}\\
  \textit{Tianjin University}\\
  Tianjin, China\\
  \texttt{mq12335@tju.edu.cn}
}
&
\parbox{7cm}{
  \centering
  \textbf{Yuwei Guo}\\
  \textit{College of Intelligence and Computing}\\
  \textit{Tianjin University}\\
  Tianjin, China\\
  \texttt{2024244171@tju.edu.cn}
}
\end{tabular}
}

\maketitle

\begingroup
\renewcommand\thefootnote{}\footnotetext{\textsuperscript{*} Ming Zhang and Yiling Xuan are co-first authors and contributed equally to this work.}
\endgroup

\begin{abstract}
Many real-world systems, such as transportation systems, ecological systems, and Internet systems, 
are complex systems. As an important tool for studying complex systems, computational experiments 
can map them into artificial society models that are computable and reproducible within computers, 
thereby providing digital and computational methods for quantitative analysis. In current research, 
the construction of individual agent models often ignores the long-term accumulative effect of 
memory mechanisms in the development process of agents, which to some extent causes the constructed 
models to deviate from the real characteristics of real-world systems. To address this challenge, 
this paper proposes an individual agent model based on a memory-learning collaboration mechanism, 
which implements hierarchical modeling of the memory mechanism and a multi-indicator evaluation 
mechanism. Through hierarchical modeling of the individual memory repository, the group memory 
repository, and the memory buffer pool, memory can be effectively managed, and knowledge sharing 
and dissemination between individuals and groups can be promoted. At the same time, the 
multi-indicator evaluation mechanism enables dynamic evaluation of memory information, allowing 
dynamic updates of information in the memory set and promoting collaborative decision-making between 
memory and learning. Experimental results show that, compared with existing memory modeling 
methods, the agents constructed by the proposed model demonstrate better decision-making quality 
and adaptability within the system. This verifies the effectiveness of the individual agent model 
based on the memory-learning collaboration mechanism proposed in this paper in improving the quality 
of individual-level modeling in artificial society modeling and achieving anthropomorphic 
characteristics.
\end{abstract}


\section{Introduction}
In order to analyze and study the increasingly prominent characteristics of complex networks in 
complex social systems with the rapid development of the Internet, the Internet of Things, big 
data, and social media\cite{lazer2020computational, xue2022research}, it is important to note that complex social systems involve human 
and social factors, and their design, analysis, organization, control, and integration are facing 
unprecedented challenges. Studies by Dirk Helbing, Marko Jusup, and others have shown that models 
built based on complex science methods can be widely applied in the field of social dynamics to 
help prevent problems such as urban disasters, crime, infectious diseases, war, and terrorism\cite{jusup2022social,helbing2015saving}. 

Traditional experimental methods are usually based on physical entities, but this approach cannot 
be directly applied to the study of actual social systems. The main reasons include the 
following: 1) Complex social systems cannot be studied through reductionist methods, as decomposed 
systems are likely to lose their original functions and characteristics. Therefore, the system must 
be studied as a whole\cite{xue2022computational}. 2) Due to the large scale of complex social systems, conducting repeated 
experiments on real systems is economically infeasible. 3) Many complex systems involving social 
governance are legally constrained and usually cannot be tested or constructed, such as national 
security, military preparedness, and emergency response. 4) Complex social systems are highly 
related to humans, and in many cases require human participation. Testing these systems may lead to 
irreversible risks and losses, which do not meet ethical requirements\cite{xue2023computational}.

Against this background, research on complex social systems has gradually shifted toward “computational
experiments,” which enable quantitative analysis of complex systems through algorithmic and 
counterfactual methods\cite{pearl2018book}. Figure 1 shows the workflow of computational experiments. First, by 
constructing autonomous individual models and their interaction rules, a conceptual model of the 
complex social system is abstracted from a microscopic perspective. Then, by integrating complex 
system theory with computer simulation technology, a “digital twin” of the real system is cultivated 
in the information world\cite{lou2019towards}. Next, by adjusting system rules, parameters, and external intervention 
strategies, multiple computational experiments can be repeatedly conducted. Finally, based on the 
experimental results, causal relationships between intermediate variables and system emergence can be 
identified, providing a new approach for explaining, interpreting, guiding, and reshaping macro 
phenomena in the real world. At present, computational experiments have been applied in multiple 
fields, especially in scenarios with high risk, high cost, or where experiments cannot be directly 
conducted in reality, such as intelligent transportation systems\cite{li2019parallel,li2016intelligence}, war simulation systems\cite{benhassine2024advancing}, 
socio-economic systems\cite{schweizer2023pathways}, ecological systems\cite{zhang2024investigating}, physiological/pathological systems\cite{yang2025foundation,alvarez2023computational}, 
and political-ecological systems\cite{lettieri2016computational}.

\begin{figure}[htbp]
\centering
\includegraphics[width=\linewidth]{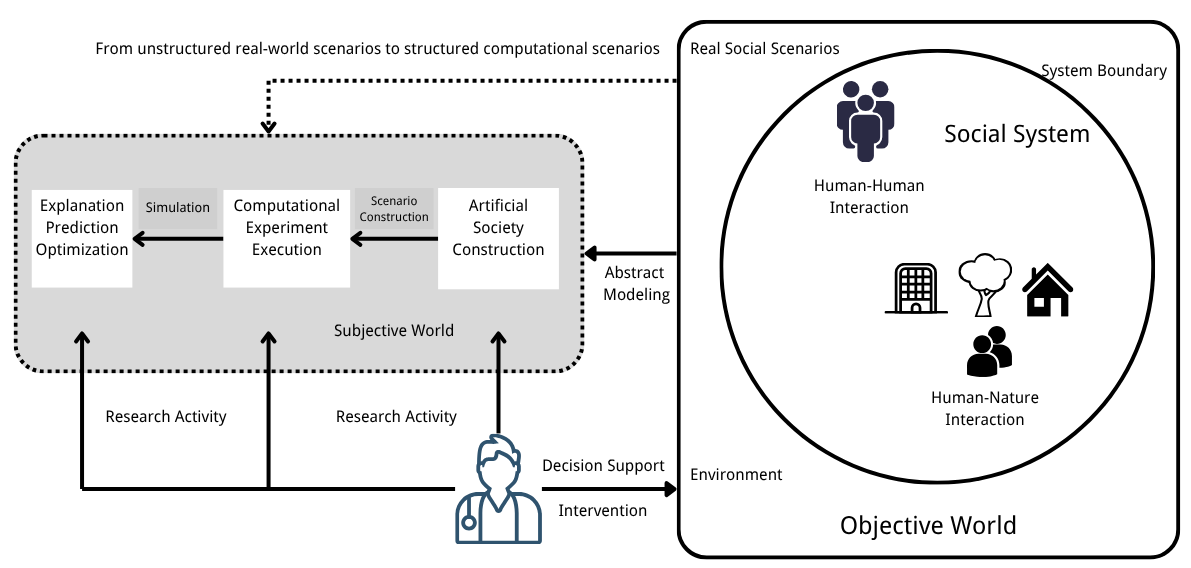}
\caption{Schematic Diagram of the Computational Experiment Method.}
\end{figure}

The technical framework of computational experiments mainly includes the following five steps: artificial 
society modeling\cite{xue2018social,zhou2022sle2,zhou2024hierarchical}, experimental system construction\cite{xue2024acomputational}, experimental design\cite{xue2024bcomputational}, experimental 
analysis\cite{xue2019analysis}, and model validation\cite{lu2021computational}. Among them, artificial society modeling serves as the foundation 
for conducting computational experiments. Only by clearly defining the structures, elements, and 
attributes required in the artificial society, and mapping the complex system into a computer-operable 
experimental system, can the simulation and observation of complex system phenomena be conducted. 
The agent-based modeling (ABM) approach\cite{ge2013framework} is an important tool for artificial society modeling. 
By effectively integrating the behavioral characteristics of social entities, structural modeling at 
the microscopic level can be achieved. In the development of complex social system research, artificial 
society modeling methods have been continuously improved and evolved, from single-agent systems to the 
integration and application of multi-agent systems\cite{deepa2016swarm}, and many achievements have been made. However, 
when using ABM to construct artificial societies, the reliability of the model is crucial. A high-quality 
model can not only more accurately reproduce the characteristics of the real world but also generate 
unexpected emergent phenomena. Therefore, in order to further improve the quality and credibility of 
the model, optimization and enhancement can be carried out at the individual level.

Individuals in artificial social systems differ from the homogeneous numerical computing units in 
traditional statistical models\cite{ye2017general}; instead, they are endowed with many heterogeneous attribute 
features and behavioral patterns. Due to this heterogeneity, modeling requires not only assigning 
unique individual characteristics, social attributes, and behavioral rules to each agent but also 
ensuring that each individual can make decisions based on the system's optimal strategy and its own 
experience. However, in current artificial society models, most individuals still behave as 
passively responsive entities driven by rules. In fact, individuals are neither completely 
unconscious nor lacking in initiative; on the contrary, they play a key role in the evolution of 
the system. In recent years, some studies have improved the modeling quality of individual agents 
by introducing learning algorithms at the individual level, thereby promoting the development and 
application of multi-agent systems. Although learning algorithms have played an important role in 
enhancing agent modeling quality, the most critical issue lies in the neglect of memory's important 
role in agent model construction. Memory enables individuals to fully utilize a broader range of 
historical information in the decision-making process, enhances their perception of long-term 
trends and environmental changes, equips agents with more comprehensive decision-making 
capabilities, improves their responsiveness to complex systems, and thereby enhances the 
adaptability and robustness of agents.

\section{Background and Motivation}

\subsection{Development process of artificial society modeling}
In the early stage of artificial society modeling development, the core modeling method was 
agent-based modeling and simulation, which is also the main research approach of complex adaptive 
systems theory\cite{wu2014agent}. Among them, the construction of individual agent models has always been the 
focus of research and serves as the foundation of model building. In the early studies, Michael 
Wooldridge and others\cite{wooldridge1995intelligent} systematically elaborated the basic theoretical framework of agents 
and proposed the BDI (Belief-Desire-Intention) model, which defines the three elements of belief, 
desire, and intention for agents. This became a classical paradigm in early individual agent 
modeling research and promoted the evolution of rule-based agents toward learning agents. 
Subsequently, Bandura and others\cite{bandura1977social} proposed the concept of observational learning. Their 
research suggested that agent learning relies not only on direct experience (such as the 
stimulus-response pattern emphasized by behaviorist psychology) but also on observing the 
behavior of others, known as observational learning (or imitation learning). They further 
proposed four key processes of observational learning: attention, retention, reproduction, 
and motivation.

With the rapid advancement of technology, especially the continuous breakthroughs in the fields 
of artificial intelligence and machine learning, individual agent models have gradually integrated 
more advanced algorithms to simulate more complex and realistic social phenomena. These modeling 
approaches have not only played an important role in theoretical research but have also been widely 
applied in several key areas such as public policy making, urban planning, economic forecasting, 
and environmental management. By simulating potential social dynamics and their possible outcomes, 
these models provide decision-makers with deeper insights and more scientific support, laying the 
foundation for solving complex problems in the real world. Ho et al.\cite{ho2016generative} introduced Generative 
Adversarial Networks (GAN) into imitation learning, using a discriminator to perform behavioral 
cloning without an explicit reward function, thereby addressing the state distribution shift 
problem in traditional behavioral cloning. Littman et al.\cite{littman1994markov} were the first to provide a rigorous 
mathematical framework for multi-agent reinforcement learning, which has become a general paradigm 
for subsequent research.

With the development of Large Language Models (LLMs), an increasing number of studies have begun 
to adopt LLMs as core controllers to construct individual agent models, aiming to achieve key 
aspects of human intelligence such as contextual learning, continual learning, and complex 
reasoning\cite{zhu2023ghost,sclar2023minding}. Artificial societies built on LLMs exhibit characteristics highly similar to 
real-world social systems at multiple levels, including organizational collaboration, rational 
competition, information dissemination, and group emergence\cite{chen2024enhancing}. For example, in the field of 
multi-agent collaboration, CAMEL\cite{li2023camel} realizes task decomposition and efficient cooperation 
through role-playing; AutoGen\cite{wu2024autogen} enables dialogue and task coordination among multiple agents 
via custom-defined agents; AgentVerse\cite{chen2023agentverse} provides functions for dynamically adjusting agent 
architectures and allows the composition of multiple agents into cooperative groups; 
Park et al.\cite{park2023generative} developed a generative agent that reproduces social behaviors similar to 
those in "The Sims" by simulating memory functions, thereby endowing agents with stronger 
interaction and adaptability. Although LLM-based agents demonstrate significant advantages in 
terms of intelligence, their application in artificial society modeling still faces certain 
limitations. A typical issue is the “hallucination” phenomenon, where LLM-agents may generate 
knowledge content beyond their role settings or task contexts. In terms of interaction mechanisms, 
LLM-agents primarily rely on natural language, which presents some inadequacies when simulating 
complex social dynamics such as cultural transmission and group behavior. In contrast, traditional 
agent-based modeling often simulates social dynamics through the transmission of state vectors 
and explicit rule settings, which can more effectively capture the interaction relationships among 
micro-level individuals. In addition, LLM-agents usually require substantial computational 
resources during execution, and when performing large-scale simulations, their efficiency is often 
lower than that of models built using learning algorithms, which are more efficient and 
controllable in operation.

In the research of individual agent modeling techniques, there are also related studies that 
incorporate memory as an important module in the model to optimize its adaptability in dynamic 
environments. Graves et al.\cite{graves2014neural} proposed a new computational model that combines neural networks 
with external memory, enabling neural networks to read from and write to external storage, thereby 
enhancing their memory and computational capabilities. Long Short-Term Memory (LSTM) networks in 
reinforcement learning also selectively memorize sequential information through gating mechanisms 
to address the long-term dependency problem of Recurrent Neural Networks (RNNs). Park et al.\cite{park2023generative} 
constructed a memory stream module in the virtual town of “Smallville” to record all experiences 
of agents in natural language and retrieve them based on three indicators: relevance, importance, 
and novelty. Piao et al.\cite{piao2025agentsociety} endowed agents with memory storage and retrieval capabilities by 
fine-tuning pretrained LLMs, allowing individuals to adjust their behavioral strategies based on 
historical experiences, such as evaluating social influence in information dissemination decisions.

\subsection{Motivation}
At present, in the research on individual agent model construction techniques, the importance 
of memory mechanisms is often overlooked. Although some studies have considered the role of 
memory in agent modeling, the modeling approaches adopted are generally simplified, mostly 
focusing on constructing either individual memory or collective memory, with relatively little 
attention paid to the relationship between the two. In particular, there is a lack of systematic 
research on how to extract useful experiences from individual memory to form collective 
experience. In LLM-based agent models, the absence of long-term memory mechanisms is especially 
prominent. Due to the lack of structured long-term memory modules, such models face significant 
challenges in knowledge maintenance and management, especially during the knowledge updating 
process, where the problem of “catastrophic forgetting” often arises\cite{fan2022minedojo,wang2024comprehensive}, that is, the 
introduction of new knowledge may unintentionally erase or overwrite previously accumulated key 
knowledge.

This study focuses on how to effectively model memory mechanisms to achieve collaborative 
decision-making between memory and learning, thereby improving the modeling quality of individual 
agents. When constructing an individual agent model, it is necessary not only to design the 
agent's basic static features (such as individual attributes and social attributes) but also to 
further define advanced behavioral rules, such as perception capability, learning mechanism, memory 
mechanism, decision-making mechanism, and behavioral patterns. These advanced behavioral rules 
are the core mechanisms through which agents continuously evolve and acquire adaptability within 
the system. However, current models ignore the important role of memory in the evolution of 
individuals, and the collaborative construction of individual agent models through memory and 
learning mechanisms provides a framework that better aligns with the decision-making characteristics 
of individuals in the real world. Existing research on memory modeling mainly focuses on the 
memory of a single agent, with limited attention to collective memory in multi-agent systems, 
neglecting the important role of collective memory in information sharing and knowledge 
accumulation. In addition, most existing memory modeling adopts static methods and lacks dynamic 
updating mechanisms, which may lead to excessively redundant memory information and affect the 
efficiency and quality of information retrieval. Therefore, constructing efficient individual 
agent models still faces many challenges, mainly including: 1) how to construct agent memory sets 
that can be dynamically updated to adapt to environmental changes and improve decision-making 
flexibility; 2) how to establish a connection mechanism between individual memory and collective 
memory to achieve knowledge and experience sharing and dissemination; 3) how to effectively 
coordinate memory and learning mechanisms to optimize the decision-making ability of agents and 
improve overall system performance.

\section{The Framework of Improved Individual Agent Model}
In artificial society modeling, the agent model at the individual level serves as the foundation 
for constructing the overall model, as it directly affects the reliability of the model and its 
ability to reproduce social phenomena. Effective agent modeling helps simulate and understand 
behavior at the microscopic level and reveals the decision-making processes and interaction 
rules of agents. The behavior of agents is influenced by multiple factors such as their 
attributes, historical background, and social environment. Therefore, simple 
assumptions (e.g., all agents are rational and follow the same rules) often fail to 
accurately reflect the complexity and dynamic changes of real societies, while the heterogeneity 
of agents and differences in their decision-making rules often lead to different global emergent 
phenomena. Hence, in artificial society modeling, the construction of individual agent models is 
crucial, as it can significantly influence agent behavior patterns and their adaptability to the 
environment.

\subsection{Overall structure of the model}
Artificial society modeling adopts a bottom-up modeling approach, in which the characteristics 
and behaviors of agents serve as the foundation. The behaviors and interactions of agents drive 
the evolution of the entire system. Based on the traditional approach of constructing individual 
agent models through learning mechanisms, this paper innovatively integrates memory mechanisms 
with learning mechanisms to further optimize the modeling technique of individual agents, enabling 
them to continuously optimize decision-making by combining contextual knowledge in dynamically 
changing complex environments, thereby exhibiting anthropomorphic characteristics.

In the modeling process, how to select and abstract individuals from the social system into agents 
in a complex system involves two key issues that need to be carefully considered: 1) the abstraction 
granularity of microscopic individuals: the abstracted agents need to strike a balance between the 
level of abstraction and granularity, retaining the essential characteristics of individuals while 
removing details irrelevant to modeling. Different abstraction methods may lead to different categories 
of agents; 2) the heterogeneity of microscopic individuals: multiple types of agents may exist in the 
system, and agents can be either homogeneous or heterogeneous. Therefore, during modeling, it is 
necessary to generalize homogeneous agents and differentiate heterogeneous agents.

The abstracted agent itself is also a complex system, with its own behavioral rules and objectives, 
and it can continuously learn and adjust its behavioral rules based on changes in internal states and 
external environments. The agent gradually approaches and achieves its predefined goals through continuous 
adjustment of behavioral rules and decision-making. The overall structure of the individual agent model 
based on memory-learning collaboration proposed in this paper is shown in Fig. 2. In this structure, the 
continuous flow of information organically connects each component to form a unified whole. The specific 
expression of the agent model is given in (\ref{agent_model}).
\begin{equation}
\mathrm{Agent} = \langle R, S_t, E_t, Y_t, A_t, M_t, N \rangle
\label{agent_model}
\end{equation}
where $R$ represents the static characteristics of the agent, which do not change over time, such as movement 
speed, gender, etc.; $S_t$ represents the dynamic characteristics of the agent, which vary over time, such as 
the role or age of the agent; $E_t$ is the set of external events perceived by the agent, which may come from 
the environment or other agents and, once received as observed information, may influence the agent’s behaviors 
and decisions; $Y_t$ is the decision-making mechanism adopted by the agent in response to external perceptions, 
stimuli, or during interactions with other agents; $A_t$ is the set of agent behaviors, including behaviors 
taken under external stimuli as well as spontaneously generated behaviors; $M_t$ is the set of memories accumulated 
by the agent, including both long-term and short-term memory; $N$ represents the constraints imposed on the agent, 
including environmental constraints and goal constraints.

In complex systems, once individuals are abstracted and the structure of the agent is defined, 
the next step is to consider how to enable the agent to adapt to continuously changing environments 
through evolution and development. As experience accumulates, agents continuously adjust their 
behaviors. According to the level of agent awareness (rationality), learning strategies can be 
divided into unconscious learning, imitation learning, and belief-based learning. In the individual 
agent model based on memory–learning collaboration proposed in this paper, the agent can 
continuously improve its adaptability by combining its own experience with the experiences of other 
agents. Based on the previously defined state, perception, memory, behavior, and decision, the 
expression of the agent behavior pattern is given in (\ref{agent_behavior}).
\begin{equation}
R\times S_{t} \times Y_{t} \times M_{t} \to A_{t} \times S_{t}
\label{agent_behavior}
\end{equation}
In this behavior pattern, the agent generates the next action plan under the guidance of 
behavioral rules by integrating its own state with perceived information and referring to its 
accumulated memory set, thereby updating its state. In this behavior set, each action $A_t$ can be 
represented by a triplet, as defined in (\ref{agent}).
\begin{equation}
A_{t} =<S_{start}^{t},O_{p},S_{end}^{t}>
\label{agent}
\end{equation}
where $S_start^t$ denotes the initial state, $S_end^t$ denotes the final state, and $O_p$ denotes the 
state transition function.

\begin{figure*}[htbp]
\centering
\includegraphics[width=\linewidth]{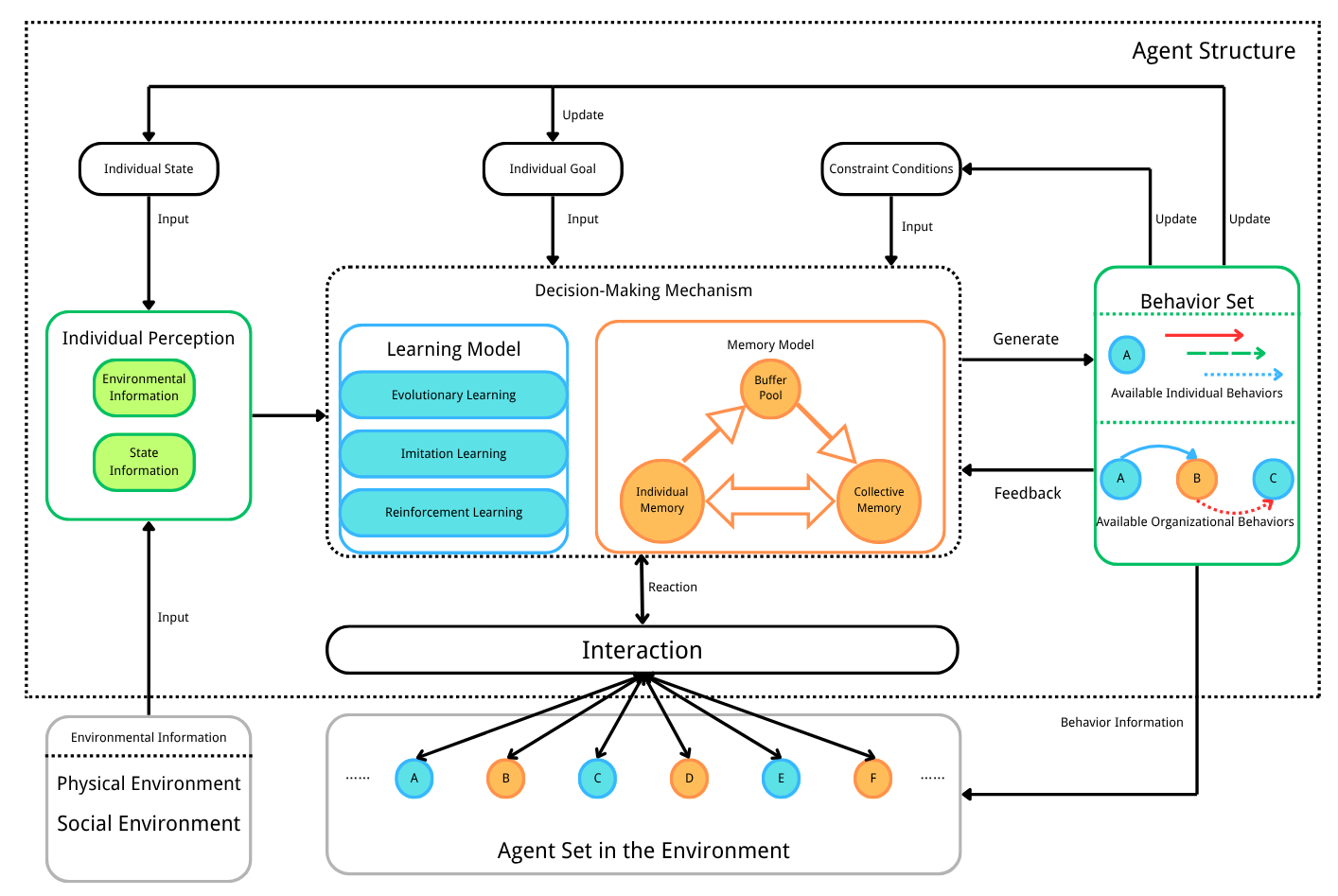}
\caption{Overall Structural Diagram of the Individual Agent Model Based on Memory-Learning Collaboration.}
\end{figure*}

\subsection{Modeling memory mechanisms}
The source of memory data is a key component in constructing the memory storage model and directly 
affects the quality of data in the memory repository. The sources of memory data can be summarized 
as the agent itself, other agents, and external knowledge bases. Agents continuously accumulate 
experience through interactions with the environment or other agents, and this experience can be 
used to optimize their behavioral strategies, thereby improving adaptability and decision-making 
capability. Meanwhile, when an information-sharing mechanism exists among agents, an agent can 
learn from the successful experiences of others, thus enhancing the overall system's coordination 
efficiency and learning ability. In addition, the rich domain knowledge and mature experience 
contained in external knowledge bases can also serve as important sources of memory data, providing 
additional informational support for the agent. These data sources interact with each other and 
collectively constitute the core content of the memory repository, providing critical support for 
the agent's learning, reasoning, and decision-making.

When constructing the memory storage model, this section divides the memory storage structure into 
three sets to better organize and manage the information in the memory model, thereby improving 
memory storage and retrieval efficiency. The three types of memory storage sets are as follows:

(1) Individual Memory Set: The individual memory set stores memory information unique to a 
single agent, including the experiences and knowledge it has accumulated through interactions 
with the environment and other agents. These memories typically involve recent events experienced 
by the agent, environmental states, decision-making processes, and the outcomes of those decisions.

(2) Collective Memory Set: The collective memory set stores information shared by all agents in 
the system, i.e., the collective knowledge base. Compared to individual memory, collective memory 
focuses more on collaboration and knowledge integration among agents, containing experiences, rules, 
and patterns contributed by different agents. This sharing mechanism promotes knowledge 
dissemination and reuse, thereby enhancing the overall intelligence level of the agent population.

(3) Memory Buffer Pool: The memory buffer pool is a dynamic storage area mainly used to 
temporarily store key information experienced by agents over short periods and, under appropriate 
conditions, select memory items to integrate into the collective memory set. Since agent systems 
generate a large amount of short-term memory during operation, and such information may not always 
have long-term value, the memory buffer pool plays a role in information filtering and management.

\begin{figure}[htbp]
\centering
\includegraphics[width=\linewidth]{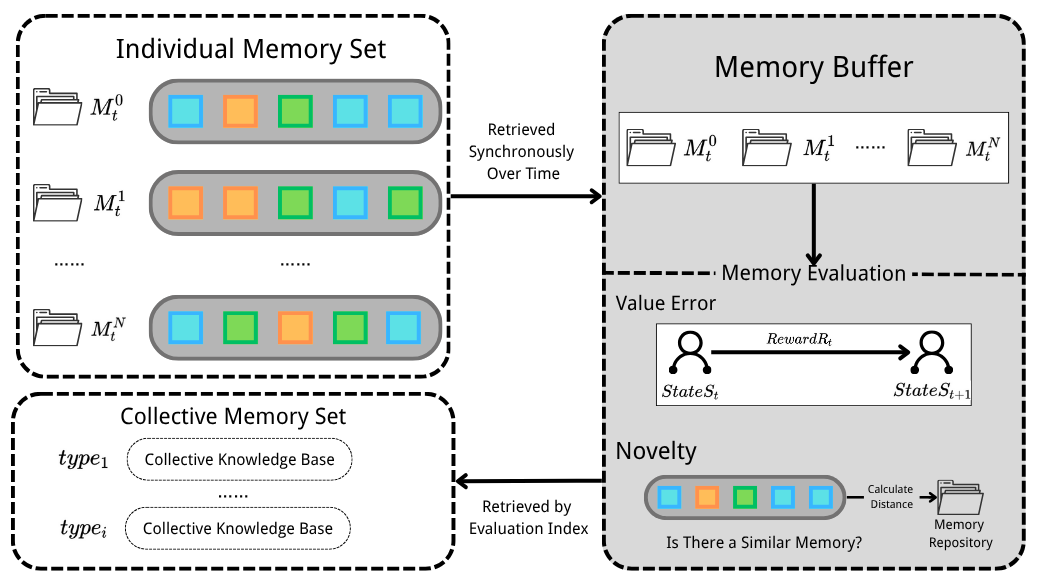}
\caption{Structural Diagram of the Memory Storage Module.}
\end{figure}

The agent's local observation information and its executed actions constitute important components 
of a memory item, and this information collectively reflects the agent's perception and 
decision-making process under a specific environmental state. After a new memory item is generated, 
it is first stored in the individual memory structure and the memory buffer pool, so that it can be 
filtered and integrated during the subsequent process of group experience dissemination and strategy 
optimization. Equations (\ref{eq4}) and (\ref{eq5}) present the mathematical description of the storage process.
\begin{equation}
M_{t}^{i} =m_{t}^{i} \cup M_{t-1}^{i}
\label{eq4}
\end{equation}
\begin{equation}
M_{t}^{buffer} =m_{t}^{i} \cup M_{t-1}^{buffer}
\label{eq5}
\end{equation}
The definition of the new memory item $m_t^i$ is given in (\ref{eq6}).
\begin{equation}
m_{t}^{i}=\left \{ type,o_{t-1}^{i} ,a_{t}^{i},o_{t}^{i}   \right \}
\label{eq6}
\end{equation}
The information in the memory item includes the memory type, the local observation $o_{t-1}^{i}$ 
perceived by the individual at the previous time step, the action $a_t^i$ executed at the current 
time step, and the local observation $o_t^i$ perceived at the current time step.

When the buffer $M_t^buffer$ is full, i.e., the memory information of all individuals at the current 
time step has been completely collected, it is necessary to construct the collective memory based 
on all the memory information in the buffer. In this process, extracting memory from the memory 
buffer pool is a key step in the dissemination of group experience. To ensure that high-value 
information can be effectively transmitted and integrated into the collective memory set, it is 
necessary to filter and evaluate the memories stored in the buffer pool, avoiding indiscriminate 
storage of all information and retaining key information as much as possible. The memory evaluation 
method comprehensively considers two aspects: value error and rarity of the memory. Equation (\ref{eq7}) defines the evaluation function.
\begin{equation}
m_{selected} =\left \{ m_{i}\in M_{t}^{buffer}\mid \left | \delta _{t}  \right | > \theta _{value} \vee R(m_{i}) > \theta _{rare}\right \}
\label{eq7}
\end{equation}
where:
\begin{itemize}
  \item $\delta_t$ is the value error of the memory item, used to measure the degree of impact that the 
  decision contained in the memory item has on the overall strategy. A larger $\left | \delta _{t}  \right | $ indicates a 
  greater influence of the memory item on the strategy, regardless of whether it represents positive 
  experience (promoting strategy optimization) or negative experience (alerting potential errors). 
  The specific calculation formula of $\delta _{t} $ is given in (\ref{eq8}):
  \begin{equation}
  \delta _{t} =\gamma V(S_{t+1})-V(S_{t} )
  \label{eq8}
  \end{equation}
  where $V$ is the state value function used to measure the value of the current state, and $\gamma$ is 
  the discount factor used to determine the importance of future value in the current decision.
  \item $R(m_i)$ is the rarity evaluation metric of the memory item, used to measure whether 
  similar experiences exist in the current system. A higher rarity indicates that the memory 
  item is more unique within the collective memory set and may contain new decision-making patterns 
  or previously unseen environmental feedback, thus providing high reference value for strategy 
  optimization. During the process of experience extraction and storage, ensuring that a certain 
  proportion of high-rarity memories are retained helps improve the diversity and adaptability of 
  strategies and prevents the model from falling into local optima. The specific calculation 
  formula is given in (\ref{eq9}):
  \begin{equation}
  R(m_{i})=\min_{m_{j}\in M_{t}^{share}} \left \| m_{i}-m_{j} \right \|
  \label{eq9}
  \end{equation}
  \item $\theta _{value} $ and $\theta _{rare} $ are the value error threshold and the rarity threshold, respectively.
\end{itemize}

To avoid problems such as increased storage pressure, decreased query efficiency, and reduced 
learning speed caused by the continuous accumulation of memory items, it is necessary to prune the 
memory set to ensure the rational use of storage space and improve the learning efficiency of the 
model. The pruning strategy filters and removes low-value or redundant memory items so that the 
collective memory can remain efficient and representative. The mathematical description of the 
pruning process is given in (\ref{eq10}):
\begin{equation}
M_{t+1}=f_{update}(M_{t},k)
\label{eq10}
\end{equation}
where $k$ is the memory length threshold, and the update process can be based on one or more 
of $\delta _{t}$ , $R(m_i)$, the number of times the memory is used, the success rate of use, and the decay 
factor. Specific pruning strategies can be adjusted according to task requirements to balance 
storage efficiency and learning performance.

\subsection{Adaptive learning mechanism}
In the individual agent model based on memory–learning collaboration constructed in this paper, 
the learning mechanism enables agents to adjust their behaviors and strategies based on external 
stimuli or internal feedback, while the memory mechanism allows agents to optimize decision-making 
by storing and recalling past experiences. However, in different learning paradigms, the 
adaptability between memory and learning, as well as their collaborative modes, vary. The following 
section analyzes the compatibility of current major learning mechanisms with memory mechanisms and 
how they work together:

(1) Evolutionary Learning

Evolutionary learning is a learning method based on natural selection, usually involving processes 
such as mutation, selection, crossover, and inheritance among agents in a population. Its goal is 
to gradually optimize the adaptability of agents through multiple generations of evolution or to 
find the strategy best suited to the current environment. The core of evolutionary learning lies in 
the intergenerational transmission of genetic information, and the evolutionary process focuses on 
improving agent adaptability through natural selection. In evolutionary learning, agent decisions 
are usually based on the current environment and genetic information (i.e., "gene" features), 
rather than on historical experience or memory. Therefore, the main driving force of the 
evolutionary learning mechanism is genetic mutation and selection rather than the accumulation of 
past experience. In view of this, the contribution of memory mechanisms to evolutionary learning is 
limited, and the combination of the two is not considered in the subsequent modeling in this paper.

(2) Imitation Learning

Imitation learning is a mechanism based on social learning, in which agents learn how to make 
decisions by observing and imitating the behaviors of others. In imitation learning, memory plays 
an important role, as agents need to store observed behavior patterns and their outcomes for 
imitation at appropriate times. Agents can adjust their imitation strategies based on previous 
observation experiences. Through memory, agents can gradually optimize behavior selection, making 
imitation learning an adaptive process. In social environments, imitation learning not only relies 
on the behavioral decisions of other agents but can also be enhanced through memory sharing among 
agents. For example, the successful experiences of certain agents may spread within the group, and 
other agents improve their adaptability by imitating these successful behaviors. This sharing and 
dissemination of memory accelerates the group learning process and enhances the overall adaptability 
of the group.

(3) Reinforcement Learning

Reinforcement learning is a learning method based on rewards and punishments, in which agents 
continuously adjust their behavioral strategies through interaction with the environment to maximize 
long-term returns. The core of reinforcement learning lies in the updating of the value function, 
that is, agents update their evaluation of behaviors based on reward signals. In this process, the 
memory mechanism plays a key role, as agents need to remember previous states, actions, and their 
outcomes to use this historical information in subsequent decisions. The memory mechanism allows 
reinforcement learning to not only rely on current reward signals but also optimize future behavior 
choices by combining past experiences. For example, in certain long-term decision-making tasks, 
agents need to remember past behaviors and understand their impact on future outcomes.

(4) Large Language Models

Agents constructed using large language models rely on learning statistical patterns and semantic 
structures in language from large-scale corpora, thereby gaining the ability to understand, 
generate, reason with, and execute tasks using natural language. During task execution, agents 
analyze and process language input to achieve task decomposition, planning, and continuous action. 
Although such agents exhibit strong flexibility and generalization capabilities, especially in 
handling complex language-dominant tasks, their underlying models rely on context windows of limited 
length for short-term information processing, and they inherently lack stable and persistent 
long-term memory mechanisms. Therefore, introducing memory mechanisms into LLM-based agent systems 
not only helps maintain consistency between task context and behavior but also provides key support 
for achieving stronger adaptability and human-like intelligence.

\subsection{Collaborative decision-making module}
The introduction of a memory mechanism enables agents to store, retrieve, and utilize historical 
experiences, thereby improving the rationality and stability of decision-making. However, relying 
solely on memory may cause agents to become overly dependent on past experiences, limiting their 
ability to adapt to new environments; whereas relying only on learning may lead to slower 
convergence and insufficient utilization of accumulated knowledge. This section focuses on the 
collaborative decision-making mechanism between memory and learning, specifically addressing how to 
extract valuable information from the memory pool and how to dynamically adjust strategies by 
integrating learning mechanisms, enabling agents to both draw on historical experience and flexibly 
adapt to new environments, thus achieving better decision-making capabilities in complex 
environments. Within this framework, the learning mechanism is responsible for generating optimal 
decisions under the current state, while the collective memory model extracts historical experience 
from the group to assist individual decision-making. Based on this idea, this section proposes a 
memory–learning collaborative decision-making model, and the specific process of the model is shown 
in Fig. 4.

\begin{figure*}[htbp]
\centering
\includegraphics[width=\linewidth]{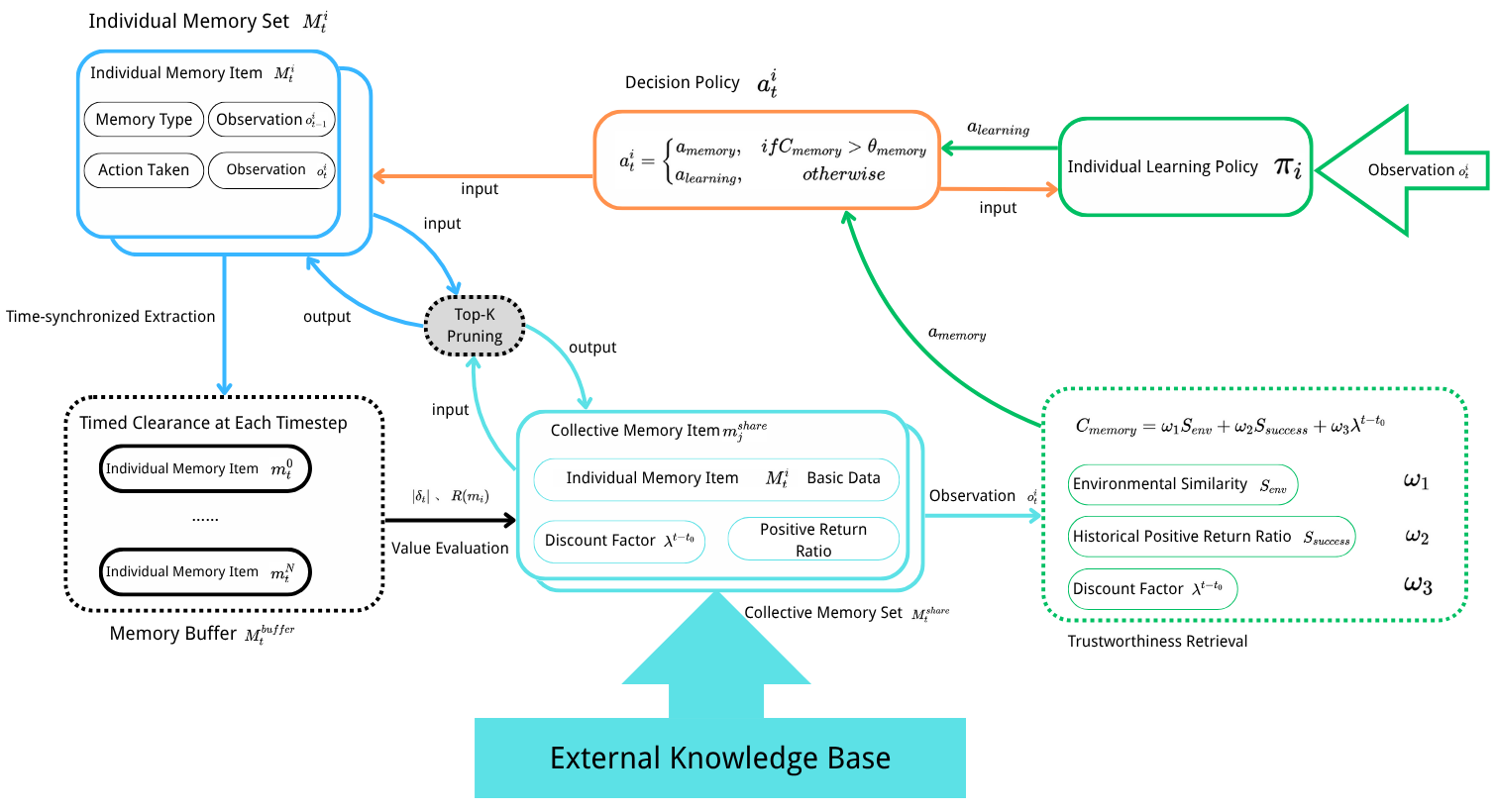}
\caption{Diagram of the Agent Decision-Making Process Based on Memory-Learning Collaboration.}
\end{figure*}

When generating the final decision $a_t^i$, the agent integrates the collective memory $M_t^share$ and 
its own decision-making mechanism to fully utilize both historical experience and individual 
learning capability. The collective memory $M_t^share$, serving as a source of global information, 
aggregates the experiences of multiple agents at different time steps and provides global reference. 
The specific mathematical description of the decision-making process is given in (\ref{eq11}):
\begin{equation}
a_{t}^{i}=\left\{\begin{matrix}a_{memory},
  & if C_{memory}> \theta _{memory}  \\a_{learning},
  &otherwise
\end{matrix}\right.
\label{eq11}
\end{equation}
where $C_memory$ is the credibility of memory experience, which is an important indicator to assess 
whether the stored memory is applicable to the current decision. Since the effectiveness of 
historical experience changes over time and is influenced by environmental similarity and the 
payoff of past decisions, it is necessary to filter and weight different memory items when using 
collective memory for decision-making, to ensure that the experiences referenced by the agent are 
the most valuable. The calculation of $C_memory$ is mainly based on three core elements, and its 
specific formula is given in (\ref{eq12}):
\begin{equation}
C_{memory} =\omega _{1} S_{env} +\omega _{2} S_{success} +\omega _{3}\lambda^{t-t_{0} }
\label{eq12}
\end{equation}
where:
\begin{itemize}
  \item $S_env$ (environmental similarity): measures the degree of matching between the current state and historical 
  memory, with a value range of [0,1]. The calculation of environmental similarity can adopt methods such as cosine 
  similarity or Euclidean distance to ensure that the agent refers only to historical experiences highly relevant to 
  the current environment.
  \item $S_success$ (positive reward ratio): represents the proportion of positive outcomes resulting from the same 
  decision in the past, with a range of [0,1]. This indicator reflects whether the memory item has contributed to effective 
  decisions in the past; the higher the success rate, the more valuable the experience is for the current decision.
  \item $\lambda^{t-t_{0} }$  represents the decay degree of memory, where $\lambda$ is the decay factor, and $t-t_0$ denotes the time 
  span between the current moment and when the memory was recorded. This term is used to reduce the influence of outdated 
  experiences and ensure that the agent gives priority to more recent and timely knowledge.
  \item $\omega _{1}$ , $\omega _{2}$ , and $\omega _{3}$  are weight parameters representing the weights of environmental similarity, positive reward ratio, 
  and memory decay term, respectively. They must satisfy the constraint: $\omega _{1} + \omega _{2} + \omega _{3} = 1$. These weights can be adjusted 
  according to specific tasks to balance the influence of different factors in the decision-making process. For example, in 
  rapidly changing environments, the weight $\omega _{3}$ can be increased so that the agent prefers more recent experiences; in stable 
  environments, the weights $\omega _{1}$ and $\omega _{2}$ can be increased to make full use of successful historical experiences.
\end{itemize}

After an individual makes a decision, the related information is stored in the memory buffer pool, 
which is used not only to optimize future decision-making processes but also to provide data support 
for the construction of collective memory. This mechanism enables agents to continuously optimize 
their behavior through the ongoing accumulation of experience, thereby forming a dynamic closed 
loop of learning and decision-making. Based on this, the system establishes a cyclic feedback 
mechanism between individual behavior and decision-making, as well as between individual behavior 
and learning strategies. Individual behavior is the result of the collaboration between memory and 
learning, and the generation of behavior not only updates the information in the memory set but 
also further optimizes individual decision-making. Through these two feedback mechanisms, 
individuals can enhance short-term decision-making capabilities using learning strategies and 
continuously optimize long-term decision frameworks with the help of memory mechanisms.

\section{Case Study: Urban instant delivery system}
To verify the effectiveness of the individual agent modeling framework based on memory–learning 
collaboration proposed in this paper, an experimental system was established in a specific 
experimental scenario. This paper selects the on-demand delivery system in modern service systems 
as the experimental scenario, due to its high dynamism and uncertainty. Delivery tasks are usually 
accompanied by multiple constraints, requiring agents to possess real-time decision-making 
capabilities and environmental adaptability. In addition, this system naturally exhibits 
characteristics such as distributed decision-making, complex interactions, and multi-task 
processing, which can fully simulate the behavioral patterns of individual agents in complex 
social environments. This complexity helps to deeply analyze the performance of agents in 
handling multiple tasks and constraints under the support of memory and learning mechanisms, 
thereby evaluating the advantages and disadvantages of different agent modeling strategies. 
Meanwhile, the on-demand delivery system, as a typical real-world application scenario, is 
representative, and its modeling approach and experimental results can be extended and applied 
to other multi-agent systems with similar characteristics, such as urban traffic scheduling, 
emergency response, and logistics management.

\subsection{Construction of an Artificial Society}
Through in-depth analysis of the on-demand delivery system, this paper abstracts the active entities 
in the system into three types of agents: Order Agent, Platform Agent, and Delivery Agent. The Order 
Agent is the core resource of the system and is managed and scheduled by the Platform Agent. As 
the central link between Order Agents and Delivery Agents, the Platform Agent is responsible for 
task allocation and information transmission. The Delivery Agent is responsible for actual delivery 
tasks, with its main activities including path planning, task execution, self-evolution, and 
interactive feedback.

(1) Order Agent
In this system, the order is the core resource and serves as a critical hub for coordinating 
operations across all components. Each order is not only a manifestation of user demand but also 
contains all key information required for delivery, such as delivery address, order value, and 
order status. The generation of an order marks the starting point of the delivery process and 
directly influences the scheduling and operational state of the Delivery Agent. It also determines 
the workflows and task distribution among various types of agents in the system, including Platform 
Agents and Delivery Agents. The attribute definition of the Order Agent is as follows:
\begin{multline*}
\text{Agent}_{\text{user}} = \langle \text{ID},\ (x_{\text{start}}, y_{\text{start}}),\ (x_{\text{end}}, y_{\text{end}}),\ \text{Status}, \\
\text{Value},\ T_{\text{start}},\ T_{\text{alive}},\ G_n() \rangle
\end{multline*}
\begin{itemize}
  \item $ID$: the identifier of the order, which uniquely distinguishes each order.
  \item $(x_{start},y_{start} )$: the starting location of the order, with the 
  coordinates constrained within the valid range of the system map.
  \item $(x_{end},y_{end} )$: the destination location of the order, with the 
  coordinates constrained within the valid range of the system map.
  \item $Status$: the status of the order, indicating its current state in 
  the system, which can be categorized as alive, dead, or captured.
  \item $Value$: the value of the order, which is related to the distance 
  between the starting and destination points as well as the order generation time.
  \item $T_{start}$: the time the order is generated, represented in 24-hour format.
  \item $T_{alive}$:  the lifespan of the order; orders do not remain in the system 
  indefinitely and will become dead once the lifespan is exceeded.
  \item $Gn()$: the order generation function, which controls the number of orders 
  generated over time; all orders are generated by this function.
\end{itemize}

In the constructed artificial society, a 24-hour system is adopted to align the simulation 
time with real-world characteristics, with a distinction made between daytime and nighttime. 
The value of an order is mainly influenced by two factors: the distance between the order's 
starting and ending points, and the time at which the order is generated. The specific calculation 
formula for order value is given in (\ref{eq13}):
\begin{equation}
Value=distance((x_{start},y_{start}),(x_{end},y_{end}))\times \xi _{time}
\label{eq13}
\end{equation}
where $distance()$ represents the actual distance obtained from the path planning results, 
and $\xi _{time}$  is the time weight coefficient. When the order is generated between 08:00 and 22:00, 
the weight is set to 1; for other times, which are considered nighttime working hours, the weight 
is set to 1.5 in order to incentivize agents to work at night and to prevent a large number of 
nighttime orders from expiring.

\begin{figure*}[htbp]
\centering
\includegraphics[width=\linewidth]{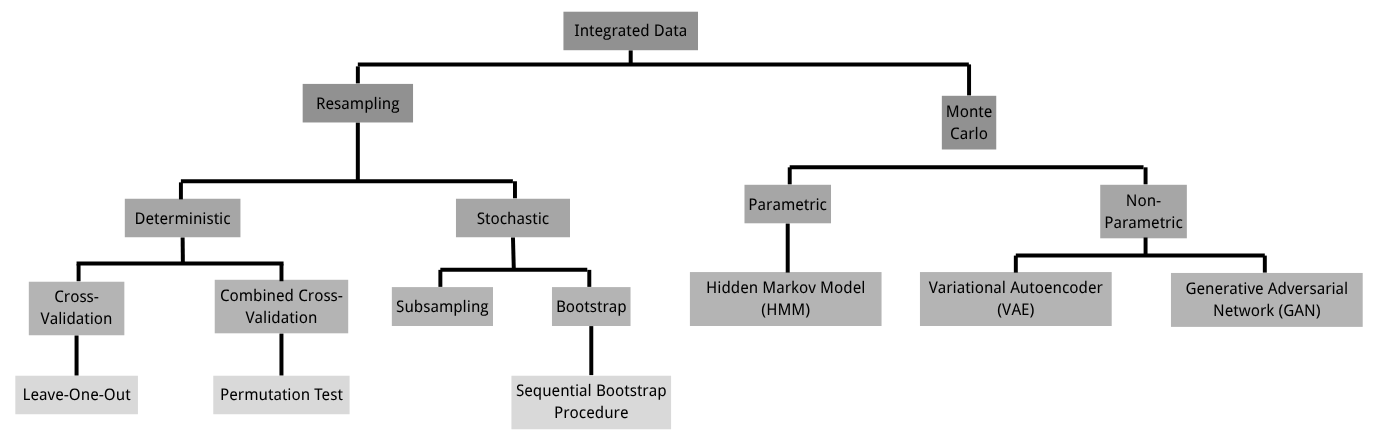}
\caption{Data Generation Methods for Computational Experiments.}
\end{figure*}

To better simulate the distribution of order generation in the real world, there are various 
methods for generating datasets in computational experiments, which can be mainly divided into 
two categories: resampling methods and Monte Carlo methods. The specific generation method is 
shown in Fig. 5. This section adopts the Monte Carlo simulation method to simulate the order 
generation process. Web crawler technology is used to collect relevant order review information 
from the ZBJ.com, and the review time is taken as the service completion time. 
The obtained data is cleaned and categorized, and the order generation pattern is fitted according 
to the chronological order. A fifth-order Gaussian function is used as the order generation 
function, and the specific function is given in (\ref{eq14}):
\begin{equation}
Gn(x)=\sum_{i=1}^{5} a_{i}e^{(-(\frac{x-b_{i}}{c_{i}})^{2} )}
\label{eq14}
\end{equation}
The values of $a_i$, $b_i$, and $c_i$ are shown in Table 1.
\begin{table}[htbp]
\caption{Order Generation Function Parameters}
\centering
\begin{tabular}{c|ccccc}
\hline
 & 1 & 2 & 3 & 4 & 5 \\
\hline
$a_i$ & 314.2 & 188.3 & 95.56 & 22.9 & 48.67 \\
$b_i$ & 172.5 & 281.5 & 315.5 & 228.9 & 267.1 \\
$c_i$ & 4.645 & 1.559 & 10.69 & 167.7 & 13.1 \\
\hline
\end{tabular}
\label{tab:order_parameters}
\end{table}

(2) Platform Agent
As the coordinator and manager, the Platform Agent mainly undertakes key responsibilities such 
as resource scheduling, task allocation, and task information management. In the artificial 
society model constructed in this section, the core function of the Platform Agent is to 
effectively allocate and schedule order resources in the system. Suppose there is an order 
set $O=\left \{ o_{1},o_{2},\dots ,o_{m}    \right \} $ and a set of idle Delivery Agents $D=\left \{ d_{1},d_{2},\dots ,d_{n}    \right \} $, the 
objective of task allocation is to minimize the total system cost, which is expressed by the 
Equation (\ref{eq15}).
\begin{equation}
cost=\sum_{i=1}^{m}\sum_{j=1}^{n} c_{ij}
\label{eq15}
\end{equation}
where $c_ij$ is the cost of assigning order $o_i$ to Delivery Agent $d_j$, and the cost is measured by 
the distance from the current location of the agent to the starting point of the order. By 
minimizing $cost$, the efficiency of order delivery can be effectively improved, response time 
can be shortened, and the consumption of order resources due to timeout in the system can be 
reduced.

In addition to task allocation, another important function of the Platform Agent is order 
information management. When an order is generated, the Platform Agent needs to promptly add 
the new order to the order set so that it can be allocated in subsequent iterations; when an 
order is assigned to a specific Delivery Agent, its status must be tracked, and the order should 
be removed from the unassigned set; after the delivery is completed, the order should be removed 
from the set, and the agent that performed the task should receive an appropriate reward; if the 
order is not completed within the time limit and expires, the Platform Agent must remove the 
expired order from the order set in a timely manner to prevent it from being received by any 
Delivery Agent, thus ensuring the system operates accurately and efficiently.

(3) Delivery Agent
The Delivery Agent is the entity in the system that possesses intelligent behavior. By definition, 
the attributes of a Delivery Agent can be represented as Equation (\ref{eq16}).
\begin{equation}
Agent_{Deliver}  = <R, S_t, E_t, M_t, Y_t, A_t, N>
\label{eq16}
\end{equation}

\begin{itemize}
  \item R refers to the static attributes of the Delivery Agent, also known as characteristic attributes, 
  which describe the features that do not change over time. The main static attributes are shown 
  in Table 2.
  \begin{table}[htbp]
  \caption{Static Attributes of Delivery Agents}
  \centering
  \begin{tabular}{p{2.3cm} p{5.0cm}}
  \hline
  \textbf{Static Attribute} & \textbf{Description} \\
  \hline
  \textit{$ID$} & Unique identifier assigned to each delivery agent in the system \\
  \textit{$speed$} & Maximum distance an agent can move per step; fixed value is 10 \\
  \textit{$scope$} & Vision range; fixed maximum distance is $10 \times \textit{speed}$ \\
  \textit{$cost$} & Survival cost; fixed cost per step is 10, positively related to speed \\
  \textit{$type$} & Learning type; supports rule-based, imitation, and reinforcement learning \\
  \hline
  \end{tabular}
  \label{tab:static_attributes}
  \end{table}
  \item $S_t$ refers to the dynamic attributes of the Delivery Agent, which change continuously with 
  the agent's adaptive behavior. By adjusting these attributes, the agent can improve its 
  adaptability. The main dynamic attributes of the Delivery Agent are shown in Table 3.
  \begin{table}[htbp]
  \caption{Dynamic Attributes of Delivery Agents}
  \centering
  \begin{tabular}{p{3.0cm} p{5.5cm}}
  \hline
  \textbf{Dynamic Attribute} & \textbf{Description} \\
  \hline
  \textit{location} & Initial position where an agent starts receiving orders; randomly assigned on the map at initialization \\
  \textit{time\textsubscript{working}} & Working hours; defines time period during which the agent accepts orders; initialized as a continuous random 8-hour period \\
  \textit{region\textsubscript{working}} & Preferred working region; indicates areas where the agent prefers to receive orders during working time \\
  \textit{status} & Status of the agent: 0 for inactive, 1 for idle, 2 for delivering \\
  \textit{earning} & Accumulated revenue, increases as the agent operates over time \\
  \hline
  \end{tabular}
  \label{tab:dynamic_attributes}
  \end{table}

  \item $E_t$ refers to the perceptual attributes of the Delivery Agent, and the agent’s perception 
  ability affects its decision-making process. The agent’s perception is mainly reflected in two 
  aspects: perceiving information about other agents in the system and perceiving environmental 
  information. The main perceptual attributes are shown in Table 4.
  \begin{table}[htbp]
  \caption{Perceptual Attributes of Delivery Agents}
  \centering
  \begin{tabular}{p{3.5cm} p{5.0cm}}
  \hline
  \textbf{Perceptual Attribute} & \textbf{Description} \\
  \hline
  Perception of Other Agents & The agent can perceive other encountered agents and exchange information with them \\
  Perception of Environment & The agent can perceive environmental changes, including weather, traffic, and time \\
  Perception of Orders & The agent can perceive orders within its visual range and decide whether to accept them \\
  \hline
  \end{tabular}
  \label{tab:perceptual_attributes}
  \end{table}

  \item $M_t$ refers to the individual memory model of the Delivery Agent, in which the real events 
  experienced by the agent are stored. The individual memory model not only provides contextual 
  information for the agent but also serves as an important component in the construction of 
  collective memory. In the on-demand delivery system modeled in this section, memory types 
  include both long-term memory and short-term memory. Long-term memory represents the long-term 
  goals of the Delivery Agent and serves as the agent’s long-term pursuit during system evolution. 
  In system design, different agents can be assigned different long-term memory objectives; for 
  example, some agents may pursue maximum profit, while others may pursue maximum comfort, thereby 
  ensuring heterogeneity among agents. To explore the impact of the combination of memory and 
  learning mechanisms on individual evolution, all agents in this system are uniformly assigned 
  the long-term memory objective of maximizing profit. Short-term memory refers to the collection 
  of events experienced by the agent, which gradually decays over time and is eventually eliminated 
  through the pruning module. The main memory attributes are shown in Table 5.
  \begin{table}[htbp]
  \caption{Individual Memory Attributes of Delivery Agents}
  \centering
  \begin{tabular}{p{3.5cm} p{5.0cm}}
  \hline
  \textbf{Memory Attribute} & \textbf{Description} \\
  \hline
  \textit{memory\textsubscript{type}} & Type of memory, including long-term and short-term memory \\
  \textit{event\textsubscript{type}} & Type of events, mainly including delivery events, weather events, and traffic events \\
  \textit{memory\textsubscript{long}} & Long-term memory, uniformly set for maximizing reward \\
  \textit{memory\textsubscript{short}} & Short-term memory, records individual experiences. Data structure includes: [time, current location, current state, current perception, action, reward, and post-transition perception] \\
  $\lambda$ & Memory decay coefficient, set to 0.9. When memory decays below 0.1, it is discarded. Effective duration is 22 days \\
  \hline
  \end{tabular}
  \label{tab:memory_attributes}
  \end{table}

  At each time step, the memory items of the individual are stored in the memory buffer pool. 
  The information entries stored in the buffer pool are consistent with the individual's memory 
  entries. At the end of each time step, the memory information in the buffer pool is extracted 
  to become part of the collective memory set. The parameters and related function settings 
  involved in the extraction process are shown in Table 6. Among them, $L_best$ denotes 
  the optimal path length, $L_rest$ denotes the remaining path length, $L_past$ denotes the traversed 
  path length, and $N_{orders}(scope) $ represents the number of orders within the perception range. 
  The memory pruning process scores all memory items, and the top k memory items are retained 
  based on the scoring results. The specific scoring function is given in (\ref{eq17}).
  \begin{equation}
  score(m_{i})=\left | \delta _{t}  \right | +S_{success}+\lambda ^{t-t_{0} }
  \label{eq17}
  \end{equation}
  \begin{table}[htbp]
  \caption{Parameters for Collective Memory Construction}
  \centering
  \begin{tabular}{p{3.0cm} p{5.5cm}}
  \hline
  \textbf{Parameter/Function} & \textbf{Value or Definition} \\
  \hline
  $V$ & $V = \frac{L_{\text{best}}}{L_{\text{rest}} + L_{\text{past}}} \times (\text{status}_t - \text{status}_{t-1}) + \frac{N_{\text{orders}}(\text{scope})}{\text{scope}^2}$ \\
  $\gamma$ & $0.8$ \\
  $R$ & $R(m_i) = \min\limits_{m_j \in M_t^{\text{share}}} \left\| m_i - m_j \right\|$ \\
  $\theta_{\text{value}}$ & $0.9$ \\
  $\theta_{\text{rare}}$ & $0.6$ \\
  $k$ & $4000$ \\
  \hline
  \end{tabular}
  \label{tab:collective_memory_parameters}
  \end{table}

  \item $Y_t$ refers to the decision-making mechanism of the Delivery Agent, which is formed by 
  the combined effect of the learning-based decision and the memory-based decision. In this 
  system, the modeling of the individual learning mechanism includes three approaches: 
  rule-based, imitation learning, and reinforcement learning. Under the rule-based mechanism, 
  the Delivery Agent does not possess learning ability, and its dynamic attributes remain 
  unchanged during the system evolution process. Its main task is to complete the orders 
  assigned by the Platform Agent and perform random walks in the environment. Imitation 
  learning adopts the Behavior Cloning algorithm, while reinforcement learning adopts the 
  Q-Learning algorithm.

  The credibility of memory determines the manner in which behaviors are adopted. The parameter 
  $\theta _{memory}$  is initialized to 0.7. When the credibility of a memory exceeds this threshold, 
  the associated behavior can be selected. The function for memory credibility is defined as 
  (\ref{eq18}):
  \begin{equation}
  C_{memory}=0.6S_{env}+0.2S_{success}+0.2\lambda ^{t-t_{0}}
  \label{eq18}
  \end{equation}
  \item $A_t$ refers to the behavioral attributes of the Delivery Agent, which determine the 
  various actions the agent can perform during the system evolution process. In this system, 
  the Delivery Agent is mainly assigned the following behavioral attributes: 1) Order acceptance: 
  accepting delivery orders and completing the pickup and delivery process according to the 
  specified route; 2) Voluntary rest: the Delivery Agent can autonomously choose when to start 
  accepting and delivering orders, but the total working time must not exceed one-third of an 
  evolution day; 3) Movement direction: when the agent is delivering an order or roaming the 
  system without an order, it will determine its direction of movement based on its memory and 
  decision-making mechanism to ensure timely delivery or to reach areas where orders are likely 
  to appear; 4) Order acceptance location: the Delivery Agent can freely choose its starting 
  location for accepting orders after resting, in order to receive new orders as quickly as 
  possible.
  \item $N$ represents the constraints imposed on the Delivery Agent, which may originate from 
  the environment or from other agents. Currently, the system imposes the following constraints 
  on Delivery Agents: 1) Reachable area: the movement range of the Delivery Agent is limited by 
  the boundaries of the environment and the agent is allowed to travel only on designated roads; 
  2) Working time: the maximum working duration of the agent must not exceed one-third of a 
  natural day during system evolution; 3) Order reception visibility constraint: the Delivery 
  Agent can only accept orders within its perception range, and orders beyond this range are 
  considered invalid; 4) Imitation visibility constraint: the target of imitation must be located 
  at the same position as the Delivery Agent to enable information exchange, otherwise imitation 
  cannot occur; 5) Dynamic attribute change constraint: the dynamic attributes of the Delivery 
  Agent cannot be changed frequently and may be adjusted at most twice within a natural day of 
  evolution. These constraints ensure that the behavior of the Delivery Agent aligns with the 
  overall requirements of the system; 6) Congestion constraint: when a certain road segment in 
  the system becomes congested, that segment will temporarily be impassable, and the agent may 
  choose to take a detour or wait for a certain period before proceeding. 
\end{itemize}

\subsection{Experiment Evaluation of Social System}
This section aims to validate the effectiveness of the individual agent model based on 
memory-learning collaboration proposed in this paper and to answer the following two research 
questions (RQ) through experiments:

RQ1: In the study of individual agent modeling techniques, how do commonly used agents constructed 
based on learning algorithms and large language models differ in their performance within the system?

RQ2: Compared with existing memory models, how does the proposed memory model perform in enhancing 
the anthropomorphic characteristics of agents and enabling high-quality decision-making within the system?

\subsubsection{Initialization of computational experiment}
The size of the experimental environment is 786 in length and 890 in width, and the map is divided into 
four subregions, each with a length of 186 and a width of 212. The underlying modeling of the map adopts a 
two-dimensional grid representation, and all positions on the map can be represented using two-dimensional 
coordinates. The basic parameter settings of the experiment are shown in Table 7.

\begin{table}[htbp]
\caption{Experimental Configuration}
\centering
\begin{tabular}{p{3.5cm} p{5.0cm}}
\hline
\textbf{System Parameter} & \textbf{Experimental Setting} \\
\hline
Environment Size & $786 \times 890$ \\
Environment Structure & Grid structure; all positions in the environment are available and represented by 2D coordinates \\
Number of Agents & 400 \\
Time Unit & 4 minutes per step \\
Number of Generations & 21600 (equivalent to 60 days in the real system) \\
\hline
\end{tabular}
\label{tab:experiment_settings}
\end{table}

To address RQ1, this section selects several representative learning models currently used in the research 
of individual agent modeling techniques for comparison. The detailed content is as follows:
\begin{itemize}
  \item Rule-based: The behavior of rule-based agents is controlled by a set of predefined rules, 
  typically using an If-Then logical structure for decision-making. Since their behavior and decision processes 
  are explicit and predictable, they can serve as a baseline model for comparison with other learning models.
  \item  Imitation Learning\cite{pomerleau1988alvinn,ross2011reduction}: Imitation learning agents learn by mimicking the behavior of experts rather 
  than relying on reward signals or environmental feedback. This paper adopts the commonly used Dagger (Dataset Aggregation) 
  algorithm in the field of imitation learning, which effectively addresses the distributional shift problem present in 
  traditional imitation learning.
  \item Reinforcement Learning\cite{watkins1989learning}: Reinforcement learning agents learn how to act through interaction with the 
  environment to maximize long-term cumulative rewards. This section uses the Q-Learning algorithm based on value iteration.
  \item Large Language Model\cite{chen2023agentverse}: Large language model agents serve as core controllers of the agent model by understanding 
  natural language input and generating appropriate natural language output. This section adopts the AgentVerse framework from 
  existing research, with the underlying model being qwen-max-0919 provided by Alibaba Cloud.
\end{itemize}

To address RQ2, this section selects several representative memory models currently used in the research of individual agent 
modeling techniques for comparison. The detailed content is as follows:
\begin{itemize}
  \item Individual Episodic Memory Model\cite{chen2024enhancing}: In such models, each agent maintains its own memory information. During the 
  decision-making process, the agent retrieves memory based on relevance, importance, and recency, and selects partially 
  related observations as guidance to adjust its response to the current situation.
  \item Collective Experience Replay Pool Model\cite{mei2023mac}: In this type of model, the memory information of all agents is 
  stored in a replay pool. During the decision-making process, the agent retrieves relevant information from the experience 
  replay pool as guidance to adjust its behavioral strategy.
\end{itemize}

\subsubsection{Case 1: The impact of different learning mechanisms on the adaptability of agents}
The experiment compares four commonly used learning mechanisms in individual agent modeling: rule-based, imitation learning, 
reinforcement learning, and large language models. By deploying these agents in the constructed on-demand delivery system 
and allowing them to evolve continuously within the system, the aim is to explore the impact of different learning mechanisms 
on agent adaptability. Next, this section will analyze the experimental results in detail from both the individual and system 
levels. Figure 6 shows the specific performance of delivery agents under different learning mechanisms and their impact on 
system performance.

\begin{figure}[htbp]
\centering
\includegraphics[width=\linewidth]{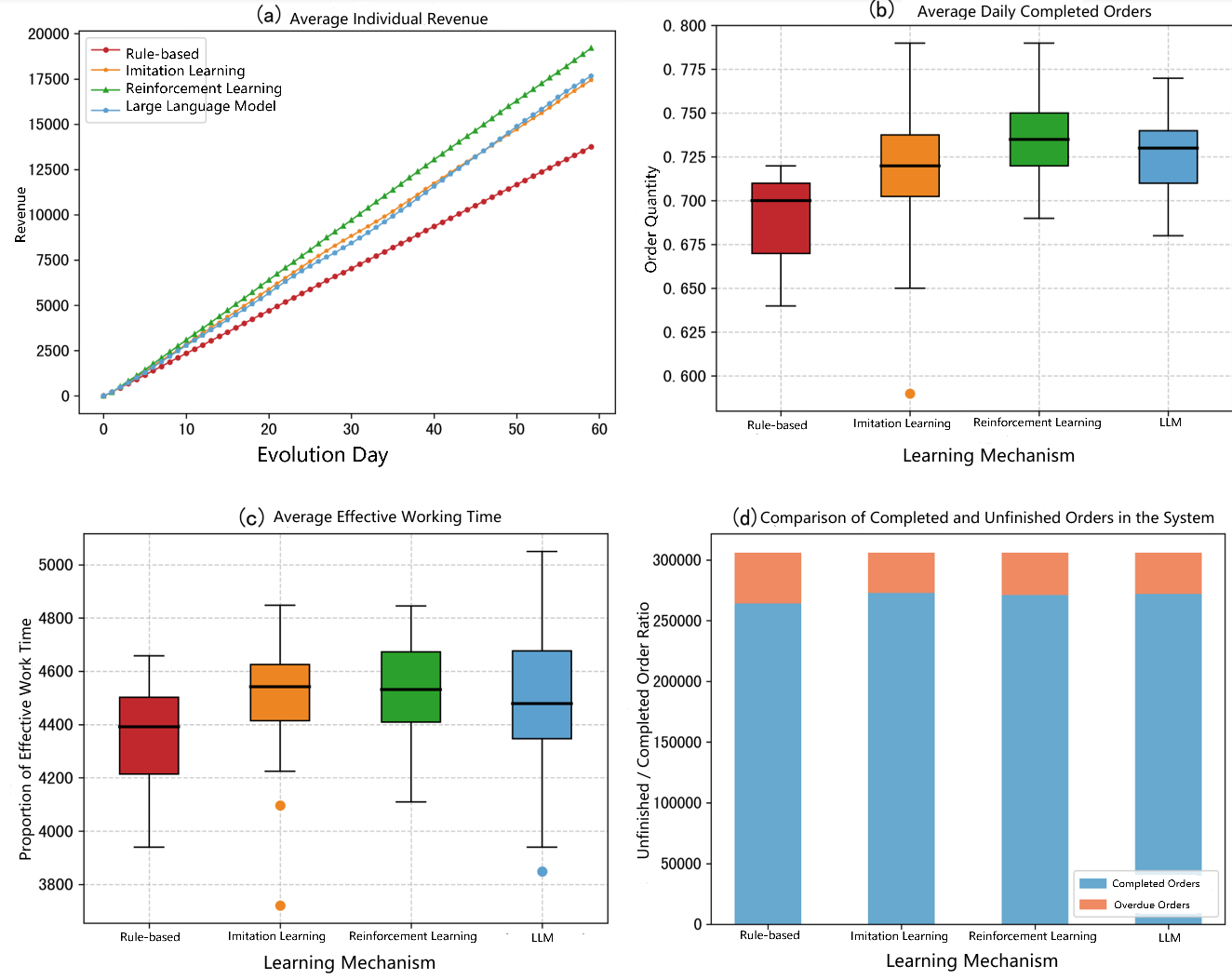}
\caption{Performance and System Impact of Delivery Agents under Different Learning Mechanisms.}
\end{figure}

From the macro-level analysis shown in Fig. 6(a-b), it can be observed that Delivery Agents constructed using learning 
algorithms exhibit a performance trend of Reinforcement Learning > Imitation Learning > Rule-based in terms of average 
individual profit and average number of orders completed per day. Agents based on large language models demonstrate 
performance similar to Imitation Learning in terms of average daily completed orders, and exhibit a trend of initially 
lower but eventually surpassing average individual profit compared to Imitation Learning. This phenomenon can be further 
explained at the micro level through Fig. 6(c), which shows the differences in effective working time of agents under 
different learning mechanisms, reflecting their varying degrees of environmental adaptability. Rule-based agents lack 
learning capability and execute tasks strictly according to predefined routes, making them unable to adapt to environmental 
changes. In contrast, Imitation Learning and Reinforcement Learning agents can continuously interact with the environment, 
optimize decision-making through learning, and enhance adaptability. Imitation Learning uses the DAgger algorithm to improve 
adaptability by mimicking the decisions of superior agents. Reinforcement Learning continuously optimizes behavior by 
training the Q-table collaboratively based on the principle of reward maximization, thus adapting to the environment more 
effectively. Large language model agents make decisions through natural language prompts and improve their decision-making 
ability and adaptability through multi-round language interactions. From a system-level perspective, Fig. 6(d) shows that 
agents with learning capabilities can optimize decisions through adaptive behavior, resulting in a higher overall order 
completion rate compared to rule-based agents that lack learning ability.

The experimental results indicate that the learning mechanism plays a critical role in enhancing agent adaptability across 
various indicators, including agent profit, average number of orders completed per day, average effective working time per 
day, and system-level order completion rate. However, compared to reinforcement learning agents, imitation learning agents 
and large language model agents exhibit certain limitations in adaptability. Imitation learning agents rely on the behavioral 
patterns of expert agents and lack an active exploration mechanism, making it difficult for them to quickly adapt in dynamic 
environments. Large language model agents lack the capability for long-term dependency and multi-step reasoning when handling 
continuous decision-making and complex environmental interaction tasks. Additionally, because their decisions are generated 
through dialogue-based interaction, the experimental process is prolonged, with most of the time spent waiting for response 
results, which limits their application in large-scale simulation scenarios. Based on the above experimental analysis, this 
paper can only conclude the performance differences among agents with different learning mechanisms under the current 
task-driven experimental scenario, but cannot directly determine whether agents constructed using large language models 
are superior or inferior to those based on traditional learning algorithms. The differences in task types determine their 
respective strengths: large language model agents perform well in natural language understanding, general reasoning, zero-shot 
tasks, and transferability, whereas agents built using traditional learning algorithms are more advantageous in high-frequency 
interaction, strategy optimization, and feedback-driven tasks.

\subsubsection{Case 2: The impact of different memory models on agent adaptability}
Experiment 2 aims to validate the effectiveness of the proposed memory-learning-based individual agent modeling framework. 
The specific method involves applying different memory modeling approaches to the Delivery Agents in the system and analyzing 
their impact on agent adaptability. By assigning different memory models to the agents with various learning mechanisms from 
Experiment 1, the agents are enabled to integrate both learning and memory mechanisms during the decision-making process. The 
specific impact of different memory models on agent adaptability is analyzed by comparing the statistical indicators of 
agents' average daily profit.

\begin{figure}[htbp]
\centering
\includegraphics[width=\linewidth]{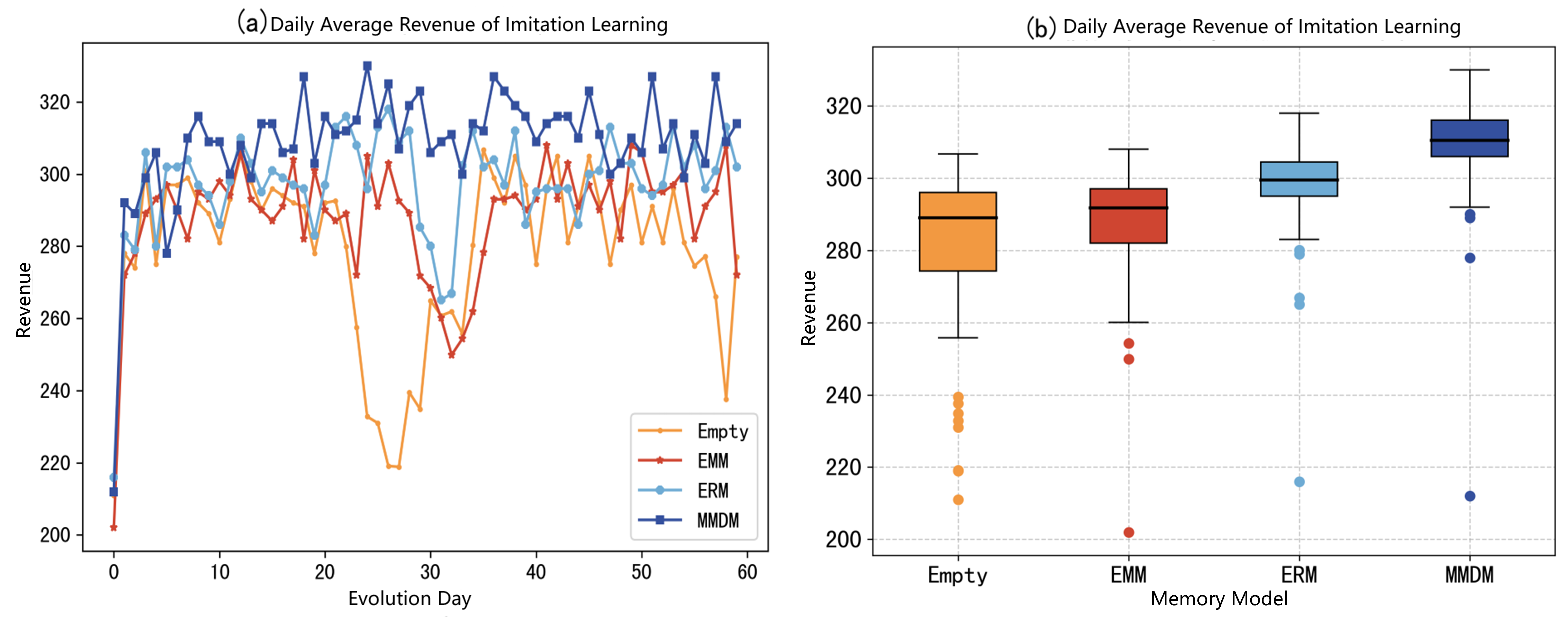}
\caption{Comparison of Daily Average Rewards of Imitation Learning Agents under Different Memory Models.}
\end{figure}

Figure 7 shows the average daily profit of Delivery Agents based on imitation learning under the assistance of different 
memory models. From the boxplot, it can be seen that memory-assisted decision-making effectively improves the agents’ average 
daily profit, among which the proposed MMDM model performs the best. In addition, the line chart reveals that in the model 
without memory, the agents' average daily profit exhibits significant fluctuations as the system evolves and only recovers 
after a long period of time. This phenomenon is mainly due to the tendency of agents in imitation learning to mimic the 
behaviors of high-reward agents during decision-making, which leads to a certain degree of behavioral clustering in the 
system. Moderate clustering can promote agents to move toward resource-dense areas, thereby improving their profit; however, 
excessive clustering may cause rapid depletion of resources in local areas, resulting in a phenomenon of “involution” among 
agents.

\begin{figure}[htbp]
\centering
\includegraphics[width=\linewidth]{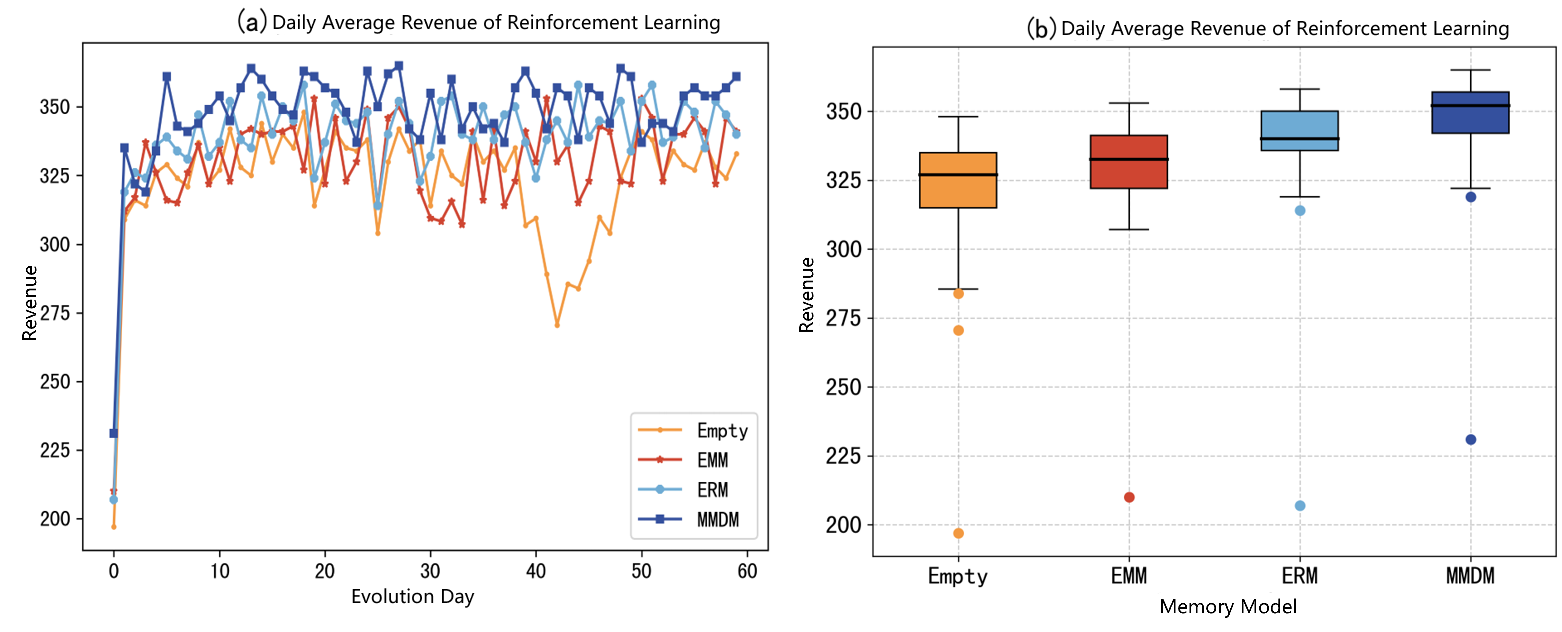}
\caption{Comparison of Daily Average Rewards of Reinforcement Learning Agents under Different Memory Models.}
\end{figure}

Figure 8 shows the average daily profit of Delivery Agents based on reinforcement learning under the assistance of different 
memory models, including a comparative analysis of line charts and boxplots. As observed from the boxplot, agent profits are 
improved under memory-assisted decision-making, with the proposed MMDM model demonstrating the best performance. The line 
chart indicates that in reinforcement learning without memory models, agent profits also exhibit significant fluctuations, 
but compared to imitation learning, the profit level recovers more quickly. This phenomenon is mainly attributed to the use 
of the Q-Learning algorithm in the system modeling process, where all agents share the same Q-table during training, leading 
to convergence in their decision-making mechanisms. However, over time, agents are able to adjust their strategies through 
continuous trial and error, gradually restoring their profit levels to the original state. In the episodic memory model, 
although memory data is limited to individual agents, there is no significant “involution” phenomenon among agents, and profits 
recover quickly after fluctuations. The line chart shows that in both the experience replay pool model and the proposed MMDM 
model, agent profits do not exhibit significant fluctuations, with only minor variations. Nevertheless, in terms of overall 
profit levels, the MMDM model performs better.

\begin{figure}[htbp]
\centering
\includegraphics[width=\linewidth]{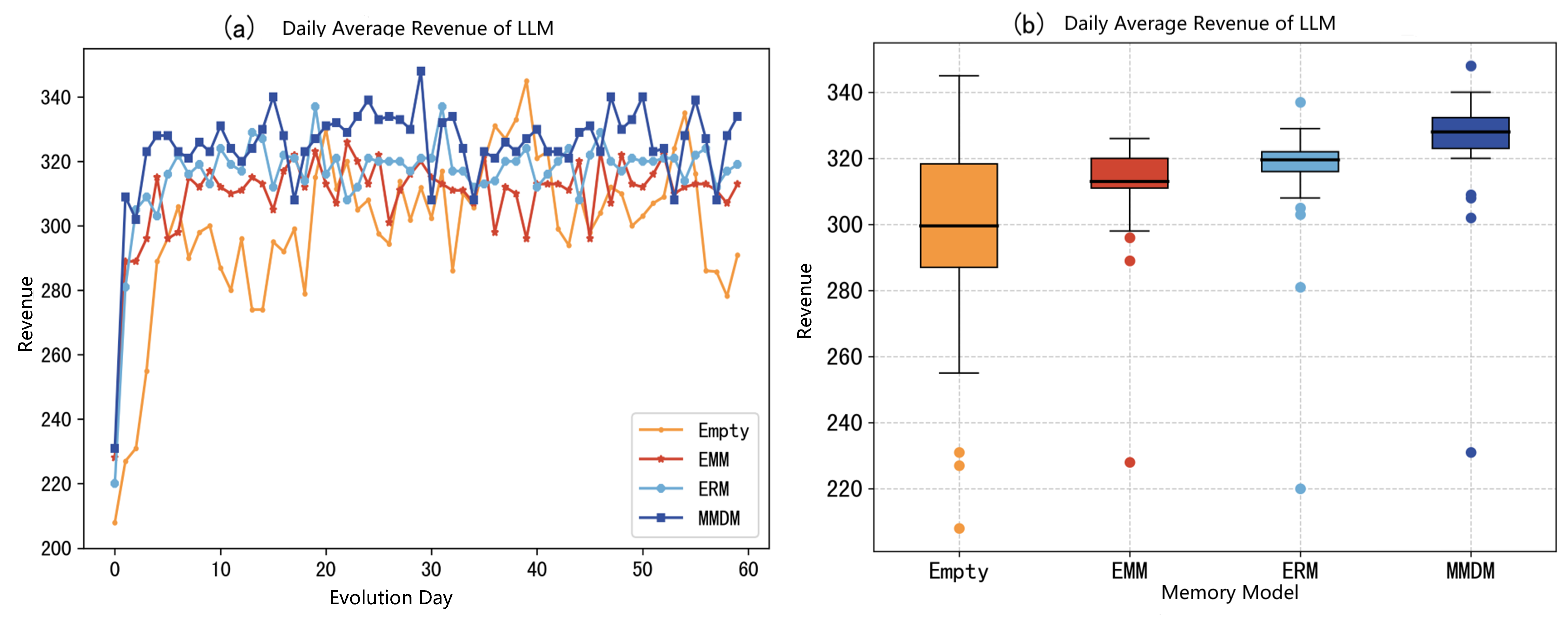}
\caption{Comparison of Daily Average Rewards of Large Language Model Agents under Different Memory Models.}
\label{fig}
\end{figure}

Figure 9 shows the average daily profit of Delivery Agents based on large language models under the assistance of different 
memory models, including a comparative analysis of line charts and boxplots. From the boxplot, it can be observed that the 
introduction of memory mechanisms improves the average daily profit of agents constructed using large language models, with 
the proposed MMDM model demonstrating the best performance. The line chart shows that in the agent model based on large 
language models without memory mechanisms, the agent's average daily profit fluctuates significantly during the evolution 
process and increases in the later stages, but no “involution” phenomenon similar to that observed in imitation learning and 
reinforcement learning models occurs. This phenomenon is mainly attributed to the fact that agents based on large language 
models without memory mechanisms rely on the model's built-in short-term context information for decision-making, resulting in 
greater volatility in the agents' performance under dynamic environments. At the same time, the short-term context allows 
agents to optimize decisions and improve adaptability to a certain extent during the evolution process. Furthermore, since each 
agent's context information is independent, excessive competition leading to “involution” does not occur.

After the introduction of the memory mechanism, the fluctuations in agent profit are significantly reduced. In the individual 
episodic memory model, agents are able to carry more comprehensive contextual information during interactions, resulting in 
better performance with less variability. In the collective memory buffer pool model, agents can utilize collective memory as 
contextual information during interactions, thus outperforming the individual episodic memory model. In the proposed MMDM model, 
high-quality memory information is ensured through the filtering of collective memory, allowing agents to carry only the most 
valuable context, thereby achieving better adaptability.

\begin{figure}[htbp]
\centering
\includegraphics[width=\linewidth]{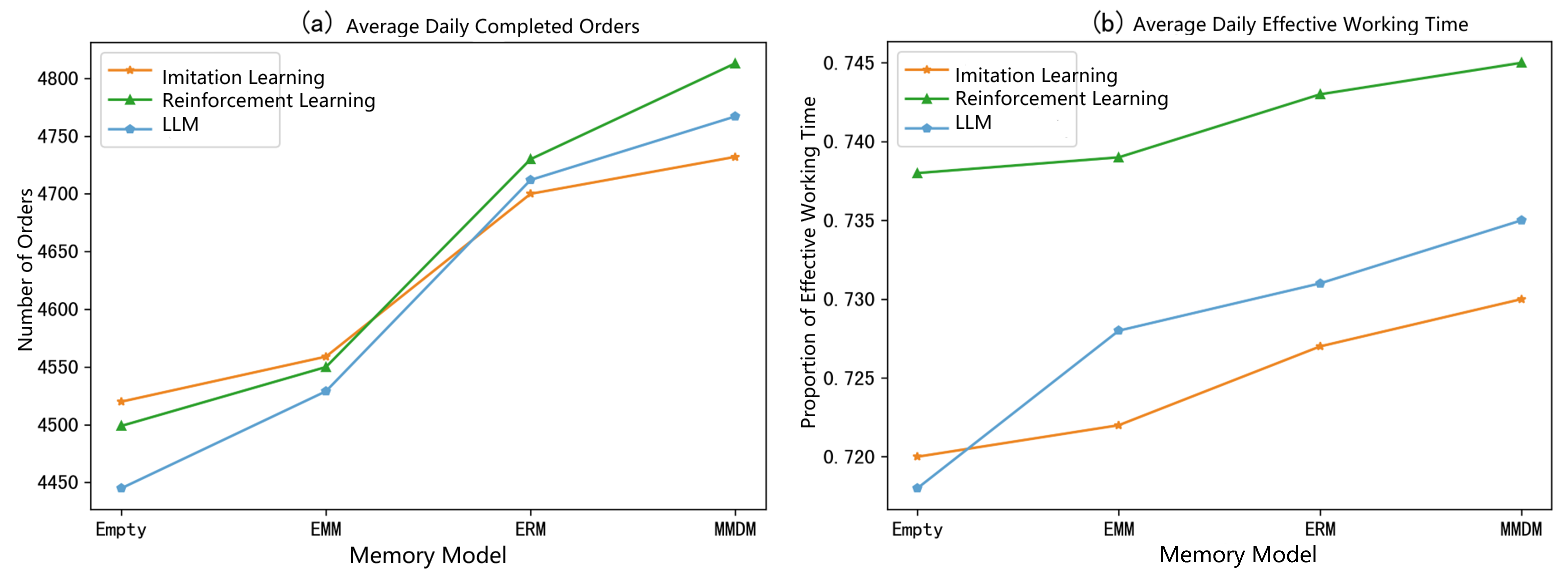}
\caption{Comparison of Agents' Daily Completed Orders and Effective Working Time under Different Decision Models.}
\end{figure}

Based on the analysis of agent profit, the internal perspective of the agent is further examined to explore the specific impact 
of different decision-making models on agent adaptability. Figure 10 presents a comparative analysis of the average number of 
orders completed per day and effective working time of agents under different decision-making models. As shown in the figure, 
the introduction of memory mechanisms improves both the agents' average daily order completion and effective working time. Among 
them, compared to the individual memory model (episodic memory model), the collective memory models (experience replay pool and 
MMDM model) have a more significant effect on enhancing agent adaptability. Overall, the integration of memory models effectively 
strengthens agent adaptability within the system, enabling agents to maintain efficient working states for longer durations and 
complete more tasks during the system's evolution process.

Through the above analysis, the effectiveness of the proposed individual agent model based on memory-learning collaboration can 
be verified. Regardless of imitation learning, reinforcement learning, or large language models, the introduction of memory 
models can significantly enhance agents' adaptability within the system. Compared with reinforcement learning, memory models 
have a more significant effect on promoting the adaptability of imitation learning agents and large language model agents. This 
is because the effectiveness of imitation learning depends on the quality of the imitation data source, while historical 
experiences in the memory set not only broaden the coverage of the imitation data source but also enhance the agent's ability 
of autonomous exploration. Under the mechanism of large language models, the introduction of the memory mechanism helps mitigate 
volatility caused by insufficient short-term information, reduces trial-and-error cost and time, and assists agents in making 
more stable and efficient decisions under similar circumstances. In contrast, reinforcement learning enhances agent adaptability 
by continuously collecting reward information through trial and error, while the introduction of memory models can reduce invalid 
exploration of agents in the system and provide valuable experiential references for the free exploration process, thereby 
accelerating the convergence process.

In addition, different types of memory models exhibit varying effects in enhancing agents' adaptability. The scenario memory 
model, as an individual memory model, primarily stores an agent's own experiences and can assist the agent in making more 
effective decisions when encountering similar situations. However, in multi-agent systems, group memory models generally perform 
better, as they can aggregate experiences from multiple agents, forming a more comprehensive knowledge base that facilitates 
collaboration and information sharing among agents, thereby improving the adaptability and efficiency of the overall system. The 
experience replay buffer is a commonly used group memory model in the current learning domain. Based on the classical experience 
replay buffer model, this paper further optimizes it and proposes the MMDM model. Experimental results validate that this model 
enables agents to select and utilize historical experiences more efficiently, thus demonstrating superior performance in complex 
environments.

Based on the overall modeling of the aforementioned artificial society, the general behaviors of agents in the system are 
abstracted into a perception module, decision-making module, behavior module, memory module, and learning module, which are 
subsequently applied to scenario construction. Through modular design, the system improves code reusability while reducing 
development costs and maintenance complexity, thereby enabling more efficient construction and optimization of artificial 
society models and avoiding redundant work of building systems from scratch each time. In addition, the system integrates a 
visual modeling tool that allows intuitive presentation of agent behavior patterns during system evolution, making it possible 
to clearly observe both micro-level agent decisions and macro-level emergent phenomena. This not only enhances the transparency 
of system operations but also facilitates deeper exploration of interaction mechanisms and evolutionary patterns in multi-agent 
systems. The system is widely applicable and can be extended beyond instant delivery systems to multi-agent application scenarios 
such as autonomous driving, robotic collaboration, and intelligent traffic scheduling.

\section{Conclusion}
In recent years, the rapid development of computational experiments has provided a new perspective for analyzing complex social 
systems, with artificial society modeling being the first step in conducting such experiments. Through artificial society 
modeling, complex social systems in the real world can be transformed into computational models, thereby enabling better 
simulation and observation of the complex evolutionary phenomena of the system. Among them, agent modeling technology serves as 
the foundation for building effective simulation systems and directly determines the model's capacity to represent and predict 
complex behaviors in real society. High-quality agent models not only reflect individual behavioral patterns more realistically 
but also reveal macro-level social phenomena emerging from multi-agent interactions. Individual agent modeling technology focuses 
on modeling the cognitive mechanisms and behavioral rules of single agents and serves as the micro-level driving force of system 
evolution. At present, the construction of individual agent models usually relies on learning algorithms or large language models 
to drive the agents' evolution in the system. However, these methods often overlook the critical role of memory mechanisms in 
individual agent models and fail to fully consider the hierarchical relationship between individual memory and group memory. 
Moreover, in many complex social systems, agents are tightly connected and exhibit strong interactions, often forming networked 
social relationship structures. Group agent modeling focuses more on the interaction relationships, coordination mechanisms, and 
network structures among multiple agents and serves as a bridge between individual behavior and emergent group phenomena. Group 
agent models enable the system's evolution to demonstrate the dynamic process from individual to group, which is key to the 
formation of macro-level social phenomena. However, existing studies mostly ignore the role of network structure in modeling or 
focus only on static network modeling, without fully considering the feedback relationship between agents and the network—i.e., 
agents are not only influenced by the network structure but also capable of influencing changes in network topology. To address 
these issues, this paper progressively optimizes existing agent models from the individual level to the group level.

The model clarifies the specific position of memory within the overall structure of the individual agent model and the direction 
of information flow among different modules during the evolution process. To effectively manage and utilize memory information, 
the model divides memory into three levels: individual memory set, memory buffer pool, and group memory set. In addition, the 
model introduces memory evaluation metrics and selectively integrates memory information into the group memory set based on these 
metrics, thereby improving the quality of group memory. To achieve dynamic updating of memory information, the model also designs 
a dynamic pruning mechanism. The model realizes the synergy between learning and memory through the coordinated decision-making 
module, jointly forming the agent's decision-making mechanism. This design not only overcomes the limitations of traditional 
models in terms of continuity and interpretability but also provides a more cognitively plausible underlying logic for simulating 
complex social phenomena. Specifically, agents in the system evolve from "behavioral imitation" to "mind reconstruction." To verify 
the effectiveness of the model, a computational experimental system is constructed in this paper, and the experimental results 
further demonstrate the validity of the model.

Based on existing research achievements, this paper further explores the research problem of improving agent modeling techniques 
at both the individual and societal levels in artificial society modeling. However, there are still some shortcomings in the 
current study that need to be addressed. Therefore, future research will focus on the following three aspects:

(1) Enhancing the quality of memory modeling to achieve agent self-reflection and adjustment: The construction of the memory 
module relies not only on computer science but also on research results from psychology, neuroscience, and other fields. By 
gaining a deeper understanding of the composition of human and animal memory, it is possible to construct more realistic memory 
modules. Historical experience is not only a core information source in the evolutionary process of agents but should also be 
used to reflect on behavior for self-optimization, thereby improving the quality of individual agent modeling and the credibility 
of artificial society modeling.

(2) Integrating large language models with social networks to jointly construct group agent models: At present, agents built on 
large language models often play the role of “respondents,” making it difficult for them to deeply understand the social structure 
in the system evolution process and lacking the ability to actively establish social connections. Future research will focus on 
how to combine social network structures with large language models to help agents better understand and integrate into social 
networks, thereby simulating more complex social phenomena.

(3) Coexistence of competition and cooperation: Through model expansion, competition and cooperation in the network are not 
isolated but intertwined to jointly form a complex interaction mechanism. By introducing the dual dimensions of competition and 
cooperation, agents can make more flexible and rational decisions in complex social scenarios, thereby effectively improving their 
adaptability and the overall system stability. In addition, such mechanisms provide new perspectives and research directions for 
exploring social behaviors, group dynamics, and collaboration issues in multi-agent systems.

\bibliographystyle{IEEEtran}
\bibliography{Reference}

\end{document}